\documentclass[aps,pre,twocolumn,groupedaddress]{revtex4-1}

\usepackage[pdftex]{graphicx}
\usepackage{color}

\begin{document}

\title{Influence of Particle Size Distribution on Random Close Packing of Spheres}

\author{Kenneth W. Desmond and Eric R. Weeks}
\affiliation{Department of Physics, Emory University, Atlanta,
Georgia 30322, USA}

\date{\today}

\begin{abstract}
The densest amorphous packing of rigid particles is known as
random close packing.  It has long been appreciated that 
higher densities are achieved by using collections of particles
with a variety of sizes.
For spheres, 
the variety of sizes is often quantified
by the polydispersity of the particle size distribution:  the
standard deviation of the radius divided by the mean radius.
Several prior studies quantified the increase of the packing
density as a function of polydispersity.  Of course, a particle
size distribution is also characterized by its skewness, kurtosis,
and higher moments, but the influence of these parameters has not
been carefully quantified before.  In this work, we numerically
generate many sphere packings with different particle radii distributions,
varying polydispersity and skewness independently of
one another. We find two significant results. First, the skewness
can have a significant effect on the packing density and in some
cases can have a larger effect than polydispersity. Second, the
packing fraction is relatively insensitive to the value of the
kurtosis.  We present a simple empirical formula for the value of
the random close packing density as a function of polydispersity
and skewness.

\end{abstract}

\pacs{}

\maketitle

\section{Introduction}

Understanding various aspects of random close packing ($rcp$) has
great scientific and industrial importance \cite{Torquato2010}
as it has been linked to a wide range of problems such
as the structure of living cells~\cite{truskett00},
liquids~\cite{Rice1944, BernelNature1960}, granular
media~\cite{Smith1929, Edwards1994, Radin2008,Jerkins2008},
emulsions~\cite{Pal2008}, glasses~\cite{Lois2009}, amorphous
solids~\cite{Zallen1983}, jamming~\cite{ohern03,ohern04}, the
viscosity of suspensions \cite{krieger59,dorr13}, and
the processing of ceramic materials~\cite{Mcgeary1961}. Random
close packing is typically defined as a collection of particles
packed into the densest possible amorphous configuration, although more
rigorous definitions are
available~\cite{TorquatoPRL2000}. Experiments have found that the
densest random packing of monodisperse spheres typically occurs
close to $\phi_{0,rcp} \sim 0.64$~\cite{BernelNature1960}, where
the density $\phi$ (or packing fraction) is defined as the ratio of
the total volume occupied by spheres to the volume of the container.

Formally, a packing consists of particles with a distribution
in radii $P(R)$.  The polydispersity is defined as
\begin{equation}
\delta =\sqrt{\langle{} \Delta R^2 \rangle{}}/\langle{}R\rangle{}. 
\end{equation} 
Here,
$\Delta R = R - \langle R \rangle$ and
the moments of $R$ (and $\Delta R$) are defined as $\langle R^n \rangle = \int
R^n P(R) dR$ [and $\langle \Delta R^n \rangle = \int
\Delta{}R^n P(R) dR$].  It has long been appreciated that packings
of spheres can have larger $rcp$ densities when $\delta > 0$
\cite{KansalChemPhys2002, Alraoush2007, Lochmann2006, brouwers06,
Okubo2004, richard01}. Prior experiments~\cite{Aim1967,
Sohn68, Dexter1972,visscher72, Weesie1994} and simulations
\cite{donev04b,donev06,hudson08,schaertl94,hermes10} have nicely
shown that as the polydispersity increases, the particles pack
to higher volume fractions because the smaller particles pack
more efficiently by either layering against larger particles
or by fitting into the voids created between neighboring
large particles~\cite{brujic09, Brouwers2006, Sohn68,
Furnas1931}. In practice, depending on the degree of the
polydispersity, the packing fraction can increase from $0.64$
for monodisperse packings to nearly $\sim 0.75$ for packings with
0.65 polydispersity~\cite{Sohn68}. For the extreme case of two
different particle sizes with a size ratio approaching infinity,
the voids between the large particles can be packed randomly with
small particles and so $\phi$ can be as large as $\phi_{0,rcp} +
(1-\phi_{0,rcp})\phi_{0,rcp} \approx 0.88$~\cite{Furnas1931, Farr2009}.

While it is intuitive that the polydispersity can affect
$\phi_{rcp}$, it is also reasonable that the shape, not just
the spread, of the distribution $P(R)$ may also influence
$\phi_{rcp}$~\cite{Torquato2010, Sohn68}. For instance, an infinite number of distributions can
have the same value of $\delta$ but yet differ in their form. One
can characterize the shape using the skewness
\begin{equation}
S = \langle{}\Delta R^3\rangle{}/\langle{} \Delta R^2 \rangle{}^{3/2}, 
\end{equation}
kurtosis
\begin{equation}
K = \langle{}\Delta R^4 \rangle{}/\langle{} \Delta R^2 \rangle{}^2,
\label{kurtosiseqn}
\end{equation}
and higher
moments. There have been prior studies that have investigated the
influence of distribution shape on the density of tightly packed
particles~\cite{Sohn68, Haughey1969, Mcgeary1961, Furnas1931,
Subbanna2002, Suzuki1985, Powell1980, Lewis1966}. Similar to the
studies on polydispersity, they find that the shape of the particle
distribution can have a profound influence on the packing density.
However, these prior studies either did not independently vary
$\delta$ and $S$ but rather conflated the influences of both,
or else used other metrics besides $\delta$ and $S$ to quantify
$P(R)$. Of the prior studies, Tickell \textit{et
al.}~\cite{Tickell1933} is the only one to report on the effects
of skewness and kurtosis for experiments carried out with sand,
finding that over a narrow range in skewness the packing
density can increase by 0.04 with no dependence on kurtosis.
However, they did not control for polydispersity, leaving
it unclear the relative importance of polydispersity and skewness.
The key unanswered question by the prior work is how the skewness
of a distribution influences $\phi_{rcp}$, and how large this effect
is relative to the effects of polydispersity.

In this paper, we address this question by numerically generating
packings with a variety of particle size distributions.  We find
that both polydispersity $\delta$ and skewness $S$ influence the maximum
random close packing volume fraction.  In particular, increasing
$\delta$ increases $\phi_{rcp}$, and for a given $\delta$,
$\phi_{rcp}$ increases linearly with increasing $S$.  As $S$ can
be negative, a negatively skewed $P(R)$ can decrease
$\phi_{rcp}$ as compared to a symmetric distribution.  We find no
universal influence of the kurtosis on our results.

\section{Protocol}

We need to generate packings with arbitrary size distributions
$P(R)$, with a goal of controlling $\delta$ and $S$
independently.  Our method for generating these
packings was previously developed in Ref.~\cite{XuPRE2005}. Briefly,
infinitesimal particles are placed randomly in a periodic container,
gradually expanded, and moved at each step to prevent particles
from overlapping. At the beginning of the simulation, particles are
assigned radii with a specific distribution and as the particles
expand they do so by a multiplicative factor such that the shape
of the radii distribution is fixed. The value of $\phi_{rcp}$
is known to be sensitive to protocol~\cite{JodreyPRA1985,
TobochnikChemPhys1988}, and it is not known if this algorithm or
any other algorithm produces rigorously defined random close packed
states~\cite{TorquatoPRL2000,XuPRE2005, Donev2004_2, OHern2004}.
Our goal is not to determine the precise value of $\phi_{rcp}$
for a given $P(R)$, but rather to empirically understand the trend
in $\phi_{rcp}$ with polydispersity and skewness.  Happily, 
our algorithm when applied to a monodisperse packing gives
$\phi_{0,rcp} \sim 0.64$, close to the experimentally found value
and in agreement with prior simulation work. In practice, the
simulation has three adjustable parameters that determine
how quickly the simulation converges to a rcp state.
These parameters are the initial packing fraction, the rate of
expansion/contraction, and a threshold on the minimum energy (see
Ref.~\cite{Desmond2009} for more details). We use the same values
as in Ref.~\cite{Desmond2009}, and we find that our algorithm
produces reproducible results and $\phi_{rcp}$ is not sensitive
to slight changes in these values.

To efficiently determine $\phi_{rcp}$ for a chosen particle
size distribution, we exploit the known finite size dependence
$\phi_{rcp}(h) = \phi_{rcp}^{\infty} - C/h$, where $h$ is the
system size, $ \phi_{rcp}^{\infty}$ is the random close packing
fraction in the limit $h \rightarrow \infty$, and $C$ is a
fitting constant~\cite{Desmond2009}. By generating many packings
with different periodic box sizes $h$, we fit $\phi_{rcp}(h)$ to determine
$\phi_{rcp}^{\infty}$ for each distribution. We generate packings
with box sizes of $\sim $10, 14, 18, and 23 mean particle diameters
in length to determine $\phi_{rcp}^{\infty}$. For the rest of the
paper, $\phi_{rcp}$ will be used to indicate $\phi_{rcp}^{\infty}$.

To control for both $\delta$ and $S$ independently, we study packings using four different distributions: binary, linear, gaussian, and lognormal. The binary and linear distributions are determined by two control parameters, allowing for us to control $\delta$ and $S$ independently, while the gaussian and lognormal distributions are determined by only one parameter, and therefore $\delta$ and $S$ can not be controlled independently. By generating many packings with different $\delta$ and $S$ using these four distributions, we can compare the results to see how sensitive $\phi_{rcp}$ is to polydispersity and skewness, but we can also compare different distributions with the same $\delta$ and $S$ to see how sensitive $\phi_{rcp}$ is to other subtle differences in the distribution shape. For all distributions, we impose $\langle R \rangle=1$.

%%%%%%%%%%%%%%%%%%%%%%%%%%%%%%%%%%%%%%
%% Table
%%%%%%%%%%%%%%%%%%%%%%%%%%%%%%%%%%%%%%
\begin{table}
\begin{center}
\begin{tabular}{lcl}
\hline
Binary & \hspace{12 pt} & \\
\hline
Function &  &$P(R) = (1-\rho)\delta(R - a) + \rho\delta(R-b)$\\
Parameters &  &Number ratio $\rho = P(b)/P(a)$ \\
 &  &Size ratio $\eta =  b/a$ \\
Constrained &  & $a = 1/(1-\rho + \eta\rho)$ \\ 
 &  & $b = \eta/(1-\rho + \eta\rho)$ \\ 
Polydispersity &  & $\delta = \left((1-\rho)(a - 1)^2 + \rho(b-1)^2\right)^{1/2}$\\
Skewness &  & $\left((1-\rho)(a - 1)^3 + \rho(b-1)^3\right)/\delta^3$ \\
Kurtosis &  & $\left((1-\rho)(a - 1)^4 + \rho(b-1)^4\right)/\delta^4$ \\
\hline
 & & \\
\hline
Linear & \hspace{12 pt} & \\
\hline
Function &  &$P(R) = AR + B$, $a \le R \le b$\\
Parameters &  & $\rho = P(b)/P(a)$ \\
 &  & $\eta =  b/a$ \\
Constrained &  & $a = 3(1+\rho)(\eta - 1)/(4-\rho)$ \\ 
 &  & $b = 3\eta(1+\rho)(\eta - 1)/(4-\rho)$ \\ 
 &  & $A = 2(\rho - 1)(4 - \rho)/(9(1+\rho)^3(\eta - 1)^4)$ \\ 
 &  & $B = 2(2-\rho)(4-\rho)/(3(1+\rho)^2(\eta - 1)^3)$ \\ 
Polydispersity &  & Solved numerically \\
Skewness &  &  Solved numerically \\
Kurtosis &  &  Solved numerically \\
\hline
 & & \\
\hline
Gaussian & \hspace{12 pt} & \\
\hline
Function &  &$P(R) = Ae^{-(R-1)^2/2\sigma^2}$\\
Parameters &  & Standard deviation $\sigma$ \\
Constrained &  & $A = 1/\sqrt{2\pi{}\sigma} $ \\ 
Polydispersity &  & $\sigma$ \\
Skewness &  & 0 \\
Kurtosis & & 0 \\
\hline
 & & \\
\hline
Lognormal & \hspace{12 pt} & \\
\hline
Function &  & $P(R) = \frac{A}{R}e^{(\ln R/\sigma + 0.5\sigma)^2)/2}$ \\
Parameters &  & Scale parameter $\sigma$ \\
Constrained &  & $A = 1/\sqrt{2\pi{}\sigma} $ \\ 
Polydispersity &  & $e^{\sigma}\sqrt{e^{\sigma^2} - 1}$ \\
Skewness &  &$\left(e^{\sigma^2} + 2\right)\sqrt{e^{\sigma^2} - 1}$ \\
Kurtosis &  & $e^{4\sigma^2} + 2e^{3\sigma^2} + 3e^{2\sigma^2} - 6 $\\
\hline
\end{tabular}
\end{center}
\caption{This table summarizes the distributions
applied in this study. The first row indicates the functional form
of the distribution. The shape of $P(R)$ is controlled by some
free parameters indicated in the next row. Each distribution is
constrained such that the mean particle radius is unity and the
probability to find any particle size is unity, which constrain
some of the coefficients in $P(R)$ to fixed values. The row titled
``Constrained" lists the fixed values of these coefficients. The
last rows list the polydispersity. skewness, and kurtosis for the
distribution.  For the binary distribution, note that $\delta(R)$
is the Dirac delta function.
For the linear distribution, analytic solutions are
unnecessarily long for the polydispersity. skewness, and kurtosis
and were computed numerically for simplicity. For the Gaussian
and lognormal distributions, a truncation was applied to ensure
that no particle radii were below 0.1. This affects the shape
slightly, and the polydispersity. skewness, and kurtosis were
computed numerically when truncation was applied.}
\label{table:Distributions}
\end{table}
%%%%%%%%%%%%%%%%%%%%%%%%%%%%%%%%%%%%%%
%%%%%%%%%%%%%%%%%%%%%%%%%%%%%%%%%%%%%%

%%%%%%%%%%%%%%%%%%%%%%%%%%%%%%%%%%%%%%
%% Figure
%%%%%%%%%%%%%%%%%%%%%%%%%%%%%%%%%%%%%%
\begin{figure}[t]
\begin{center}
\includegraphics[width=3.4in]{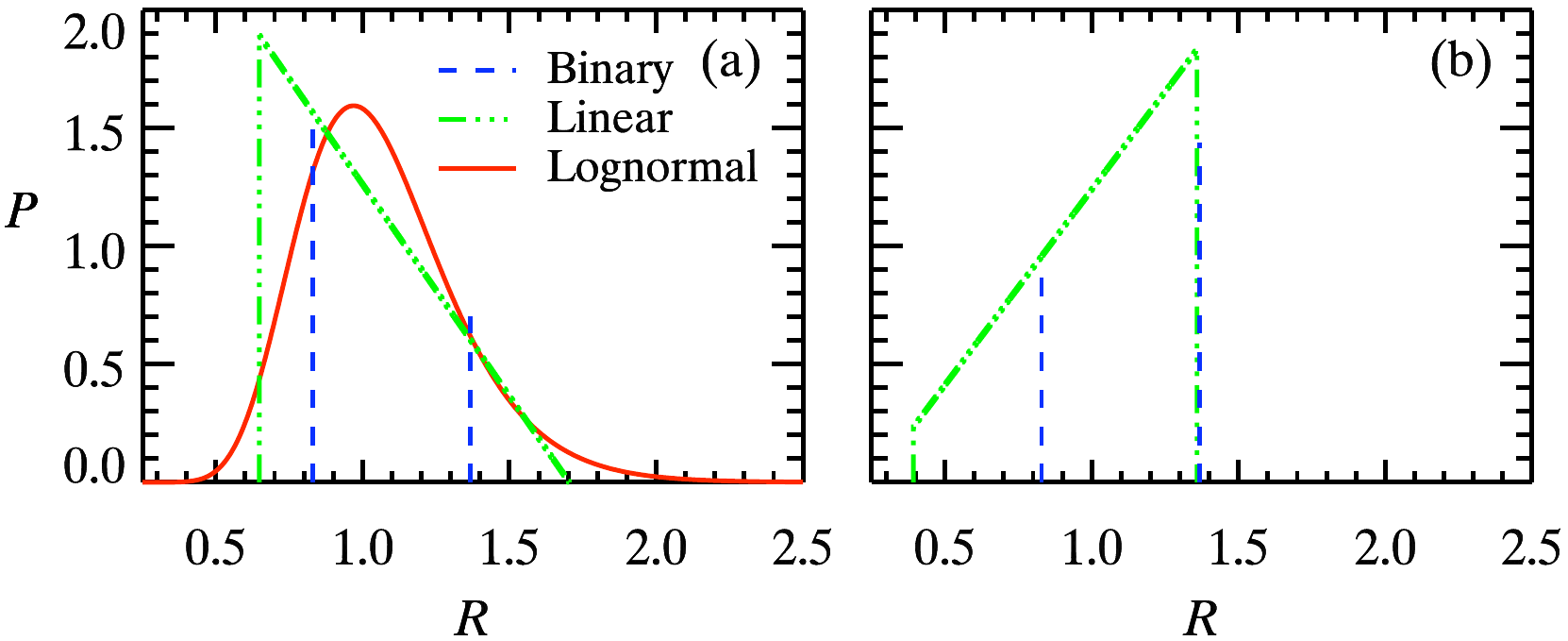}
\caption{(Color online). (a) Examples of three different particle
radii distributions with the same polydispersity of 0.25 and
nearly the positive same skewness. The binary distribution has $S
= 0.78$, the linear distribution has $S = 0.56$, and the
lognormal distribution has $S = 0.78$. (b)  Examples of two
different particle radii distributions with the same
polydispersity of 0.25 and negative skewness $S= -0.5$.}
\label{fig:CompareDistributions}
\end{center}
\end{figure}
%%%%%%%%%%%%%%%%%%%%%%%%%%%%%%%%%%%%%%
%%%%%%%%%%%%%%%%%%%%%%%%%%%%%%%%%%%%%%

More specifically, the binary distribution consists of particles
with two distinct radii.  The shape of the distribution is
determined by the size ratio and number ratio of these two particle
types.  The linear distribution is a continuous distribution of
the form $P(R) = AR + B$, where the distribution in particle size
exists over a finite range $a \le R \le b$.  Our choice of $\langle
R \rangle = 1$ and the requirement of normalization ($\int_a^b P(R)
dR = 1$) imposes two constraints on the parameters $(a,b,A,B)$.
For the two remaining degrees of freedom, we define $\eta = b/a$
and $\rho = P(a)/P(b)$.  We compute $S$ and $\delta$ for a grid
of $\eta$ and $\rho$ values, and then interpolate to find the
parameters for $P(R)$ for the desired $S$ and $\delta$ values, 
allowing us to vary them
systematically. The third distribution is a Gaussian of the form
$P(R) = A_{G}\exp(-(R - 1)^2/2\sigma^2)$, where $\sigma$ is the
standard deviation and $A_G = 1/(\sigma\sqrt{2\pi})$.  For larger $\sigma$, some of the particle radii
could be negative, which is unphysical, or very close to zero, which
may prevent generating packings within a reasonable time frame. To
avoid these issues, we truncate the Gaussian distribution such
that the smallest particle radius is no smaller than 0.1. 
The Gaussian distribution has a fixed skewness $S=0$ except for
the truncated Gaussians, which have a slight positive skewness.
The
last distribution we consider is the lognormal distribution $P(R) = A_{L} \exp(-0.5(\ln R/\sigma + 0.5\sigma)^2)/R$, where $A_L = 1/(\sigma\sqrt{2\pi})$.
Similar to the Gaussian distribution, the skewness of the
lognormal distribution is not adjustable, but is always positive
and becomes larger as $\sigma$ becomes larger. We provide a summary of the distributions in Table~\ref{table:Distributions}.

\section{Results \& Discussion}

In Fig.~\ref{fig:CompareDistributions}(a), we compare three
different distributions with polydispersity $\delta = 0.25$ and
nearly the same positive skewness $S \approx 0.75$.  We see that
the distributions are quite different, in particular in their tails.
For example, the linear distribution has many more small particles
than the other two distributions.  The lognormal distribution
has tails that include both smaller and larger particles than
the other two distributions.  It's not necessarily obvious how
the values of $\phi_{rcp}$ will be ranked for these cases. In
Fig.~\ref{fig:CompareDistributions}(b), we show two different
distributions with polydispersity $\delta= 0.25$ and skewness $S
\approx -0.5$.  As these distributions have negative skewness,
both distributions have more larger particles than smaller
particles. Once again, it's not necessarily clear how $\phi_{rcp}$
should differ between the two packings.

\begin{figure}[t]
\begin{center}
\includegraphics[width=3.4in]{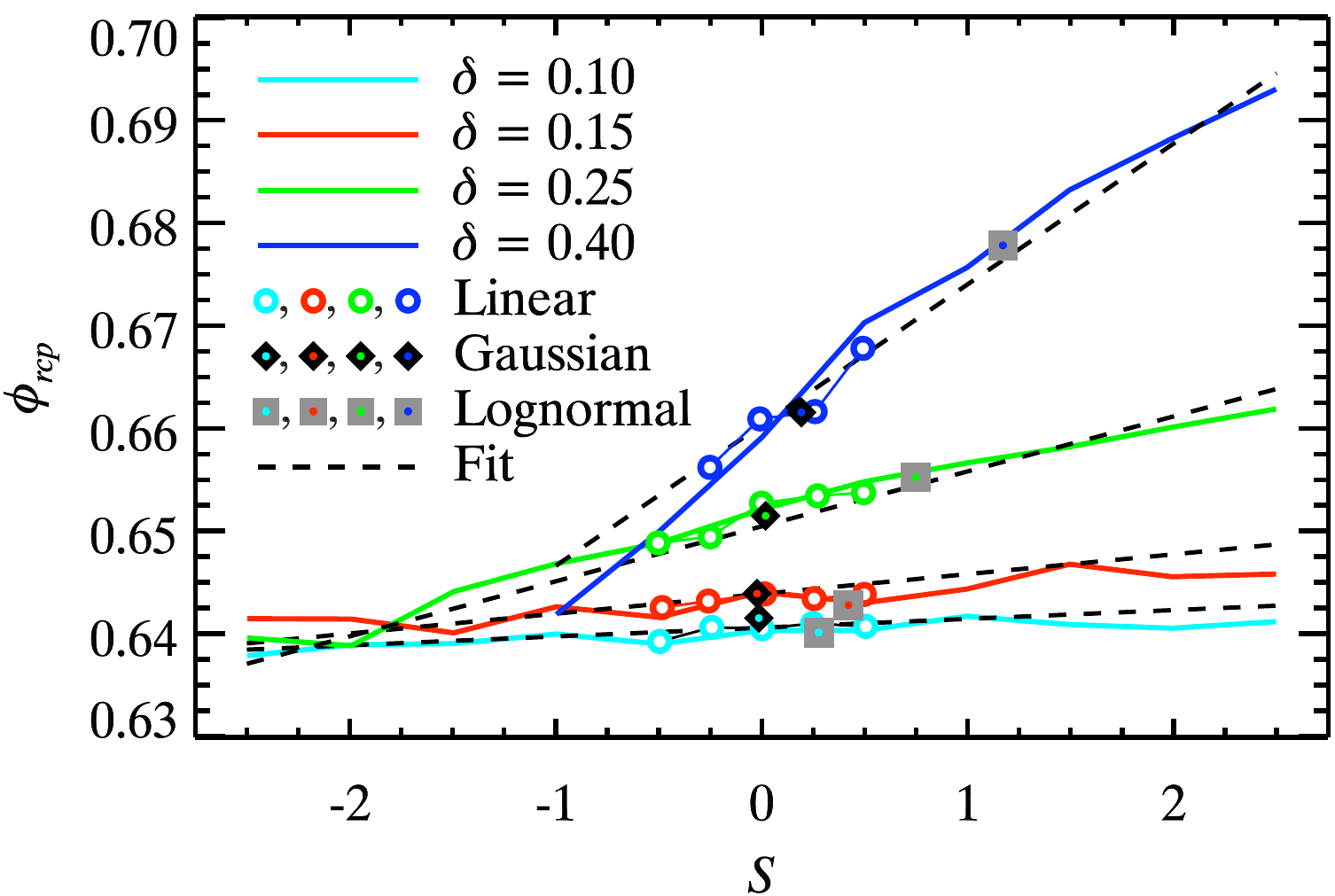}
\caption{(Color online). This figure shows how $\phi_{rcp}$ depends on particle size distribution, polydispersity $\delta$, and skewness $S$. The solid lines represent $\phi_{rcp}$ for binary packings and the symbols represent $\phi_{rcp}$ for packings with either linear, Gaussian, or lognormal particle distributions as indicated by the legend. The colors represent different polydispersities of either 0.1, 0.15, 0.25, or 0.4. The dashed lines are a fit to the data using $\phi_{rcp} = \phi_{rcp}^* + c_1\delta + c_2S\delta^2$, where $\phi_{rcp}^* = 0.634$, $c_1 = 0.0658$, and $c_2 = 0.857$.}
\label{fig:phi_rcp_data}
\end{center}
\end{figure}

After generating nearly 10,000 packings with different particle
radii distributions,
we plot $\phi_{rcp}$ as a function of skewness $S$ for all our data in 
Fig.~\ref{fig:phi_rcp_data}, with the different groups of data
(different colors)
corresponding to different polydispersity values $\delta$.
Each data point in the figure has a one to one correspondence
to the distribution type, $\delta$, and $S$. 
The symbol or line type of the data indicates
the $P(R)$ distribution type. Remarkably, the figure shows that
regardless of the type of particle radii distribution, $\phi_{rcp}$
is nearly the same for the same pairing of polydispersity and
skewness. It also shows that $\phi_{rcp}$ increases with both
increasing $\delta$ and $S$.  Strikingly, the skewness can have
an equally important effect as the polydispersity.  For example,
for $\delta=0.40$ and $S=0$, $\phi_{rcp}$ is shifted upward by $\approx
0.02$.  Fixing that value of $\delta$, changing $S$ to $\pm 1$
shifts $\phi_{rcp}$ by $\approx \pm 0.02$.  For highly polydisperse
samples, one cannot accurately know $\phi_{rcp}$ without also
knowing the skewness of the radius distribution.  For the binary
samples (solid lines in Fig.~\ref{fig:phi_rcp_data}) $S$
can be even larger in magnitude and have an even larger
influence on $\phi_{rcp}$ than $\delta$ has.

%%%%%%%%%%%%%%%%%%%%%%%%%%%%%%%%%%%%%%
%% Figure
%%%%%%%%%%%%%%%%%%%%%%%%%%%%%%%%%%%%%%
%%%%%%%%%%%%%%%%%%%%%%%%%%%%%%%%%%%%%%
%%%%%%%%%%%%%%%%%%%%%%%%%%%%%%%%%%%%%%

%%%%%%%%%%%%%%%%%%%%%%%%%%%%%%%%%%%%%%
%% Figure
%%%%%%%%%%%%%%%%%%%%%%%%%%%%%%%%%%%%%%
\begin{figure}[t]
\begin{center}
\includegraphics[width=3.4in]{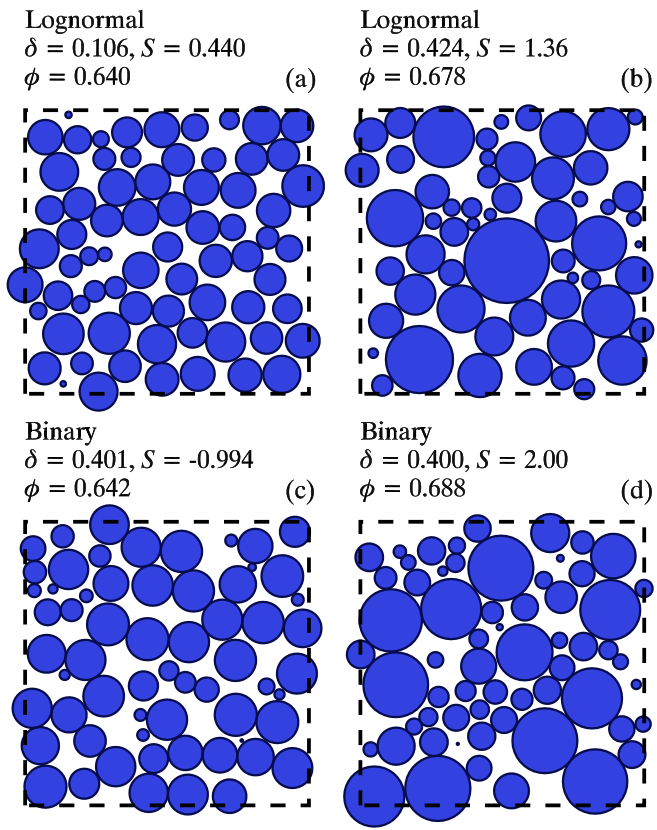}
\caption{(Color online). Each image represents a 2D slice through a 3D packing, and the dashed box is the boundary of the periodic packing. The volume fraction for each packing shown is close to the extrapolated $\phi_{rcp}$. Also, the area fraction of each 2D slice is the same as the volume fraction of the 3D packing they represent.}
\label{fig:2DSlice}
\end{center}
\end{figure}
%%%%%%%%%%%%%%%%%%%%%%%%%%%%%%%%%%%%%%
%%%%%%%%%%%%%%%%%%%%%%%%%%%%%%%%%%%%%%

The increase in $\phi_{rcp}$ with skewness is
not uniform. For negative skewness (more big particles), the
polydispersity $\delta$ does not seem to influence $\phi_{rcp}$ as
much as when the skewness is positive. This is not too surprising
since the volume of each particle grows with $R^3$. When the
total number of bigger particles is greater than the total number
of smaller particles (negative skewness), the volume occupied by
all the large particles is significantly greater than the volume
occupied by all the small particles. In effect, the big particles
pack like a low polydispersity sample and occupy the majority of
the container, while the small spheres occupy an insignificant
portion, and $\phi_{rcp}$ approaches $\phi_{0,rcp} \approx 0.64$
for a monodisperse sample.  For positive skewness (more smaller
particles), $\phi_{rcp}$ has a fairly strong dependence on $\delta$
and $S$, where $\phi_{rcp}$ increases with increasing number
of small particles.  The reason for this increase in $\phi_{rcp}$
is likely due to the small particles fitting into the spaces between
larger particles. As discussed in prior work~\cite{Brouwers2006,
Sohn68, Furnas1931}, the local porosity is smaller around
two neighboring particles of different sizes than around two
neighboring particles of the same size. This effect is greater
for larger differences in the size of two neighbors. As skewness
and polydispersity increase, both the number of small particles
present and the average size discrepancy between neighboring
particles increase, resulting in a larger $\phi_{rcp}$.

To provide a qualitative sense of the behavior,
Fig.~\ref{fig:2DSlice} shows a 2D slice through four different
3D packings. (a) and (b) are lognormal packings, where (a) is a
packing at low polydispersity and skewness and (b) is a denser
packing at a higher polydispersity and skewness. Packings (c) and
(d) are two binary packings with polydispersity 0.4, where (c) has a
large negative skewness and (d) is denser and has a large positive
skewness. As discussed above, at large $\delta$ and $S$, small
particles can either layer around larger particles and/or fit in the
voids between bigger particles. In Fig.~\ref{fig:2DSlice}(b) and (d)
we see evidence of small particles sitting in the void areas between
big particles. In Fig.~\ref{fig:2DSlice}(a) and (c), where the
skewness is lower, we see less evidence of this.  These observations
are consistent with the results of Fig.~\ref{fig:phi_rcp_data}.

Since $\phi_{rcp}$ is nearly determined by the two parameters
$\delta$ and $S$, we fit all the data to a simple equation
\begin{equation}
\phi_{rcp} = \phi_{rcp}^* + c_1\delta + c_2S\delta^2,
\label{eq:empirical}
\end{equation}
where $\phi_{rcp}^*=0.634$ is the packing fraction for a
monodisperse packing of spheres ($\delta = 0$ and $S = 0$) and
$c_1=0.0658$ and $c_2=0.857$ are empirical constants.  These fit
lines are shown as dashed lines in Fig.~\ref{fig:phi_rcp_data},
and agree reasonably well with the data.  Our fitted value of
$\phi_{rcp}^*$ is close to the experimentally accepted value of
0.637~\cite{Torquato2010, scott69}.  We also tried fits with higher
order terms in $S$ and $\delta$, but we found first order in $S$
and second order in $\delta$ reasonably fit all the data well.

There are slight differences in the $\phi_{rcp}$ values for
different distribution types for the same $\delta$ and $S$
values, seen in Fig.~\ref{fig:phi_rcp_data}.  Thus far we have
focused on $\delta$ and $S$ to characterize our distributions,
and of course the distributions differ in their higher moments.
The next quantity to consider is the kurtosis $K$ defined above
(Eqn.~\ref{kurtosiseqn}), and it might be a potential additional
parameter to explain the variations in Fig.~\ref{fig:phi_rcp_data}.
To check this, we subtract the computationally found values of
$\phi_{rcp}$ from the empirical fit (Eqn.~\ref{eq:empirical})
and plot these differences as a function of $K$ (not shown),
which shows no systematic dependence on $K$.  That is, $K$ seems
not to be a useful fit parameter for $\phi_{rcp}$.  This agrees
with the 1933 qualitative observations of Tickell \textit{et al.}
\cite{Tickell1933}.

It is worth noting that our results apply to randomly packed
objects, and one can consider other amorphous packings that are
perhaps less random.  For example, an important class of these
are random Apollonian packings (RAP) \cite{Delaney2008,reis12}.  To generate
a RAP, one first places the largest spheres, then smaller
particles are inserted into the voids between the spheres.
For example, one can fill the voids with the largest possible
spheres, \cite{Delaney2008}, which leads to a packing with a volume
fraction approaching arbitrarily close to 1.0.  In such situations,
$P(R)$ is a power law, $P(R) \sim R^{-\alpha}$ for $R > R_0$,
where $R_0$ is a cutoff.  As $R_0$ approaches 0 (an infinite
amount of iterations of the RAP protocol), the volume fraction
approaches 1.0.  Using our equations, you can see that if $R_0=0$,
$\langle{}R^n\rangle$ is infinite for $n >= \alpha-1$.  For finite
but small $R_0$, $\langle{}r^n\rangle{}$ can be quite large, and
thus the packing can have large values for $\delta$, $S$, and $K$.
Clearly in such limits the volume fraction nonetheless is 1.0 or
smaller, so our empirical formula Eqn.~\ref{eq:empirical} must
break down.  On the other hand, the RAP protocols all ensure large
volume fractions by construction -- that is, the small particles
are precisely chosen to fit into the voids between the large ones,
and $P(R)$ is determined after the fact through the algorithm.
Our computational algorithm will generally find less optimal
packings for the same $P(R)$, and so it is to be expected that
Eqn.~\ref{eq:empirical} should not apply to RAP.  

For that matter, our algorithm converges unacceptably slowly for
distributions with particle sizes varying by more than a factor
of ten between the smallest and largest sizes, preventing us from
directly testing power law packings.  We work around this by using
the numerical algorithm proposed by Farr and Groot \cite{Farr2009},
which rapidly predicts $\phi_{rcp}$ based on any $P(R)$ as input.
The predictions of their algorithm agree well with the results from
our computed 3D packings for the distributions listed in Table I.
We use their algorithm to determine $\phi_{rcp}$ for a variety of
power law distributions over the same range of $\delta$ and $S$
tested for our other distributions, and we find excellent agreement
with our empirical expression Eqn.~\ref{eq:empirical}.

\section{Conclusions}

Our data have two significant conclusions.  First, the skewness $S$
has a significant influence on $\phi_{rcp}$ for distributions with
a large polydispersity $\delta$.  Second, Eqn.~\ref{eq:empirical}
allows one to determine $\phi_{rcp}$ to within approximately $\pm
0.002$ from knowing $\delta$ and $S$, without taking
into account any other details of the shape of $P(R)$.

This collapse of $\phi_{rcp}$ values for a given $\delta$ and $S$
but different distribution shapes is intriguing, as presumably
the structures within the packings are different for different
$P(R)$.  For that matter, one can have the same $\phi_{rcp}$
value for different $\delta$ and $S$, see for example 
Fig.~\ref{fig:2DSlice}(a) and (c), and clearly these will have
different microstructures.
This might be useful for studying aspects of the jamming
transition of spherical particles.  
Many prior results show that various
properties of these systems depend on the distance to the
jamming point~\cite{Ellenbroek2006, Majmudar07,Liu2010,Desmond2013, Katgert2010_2}, where the jamming point is thought to be the
same as $\phi_{rcp}$~\cite{OHern2003,TorquatoPRL2000,Donev2004_2,OHern2004}.
One can imagine conducting experiments or
simulations to compare the properties of packings near the jamming
transition with different microstructures, but the same jamming
point.  These could be equally useful for studying the colloidal
glass transition, which may be influenced by $\phi_{rcp}$~\cite{Berthier2011_2, Liu2010, Chaudhuri2007, Hunter2012, Marcotte2013, Zhang2009_2}.
Such experiments may provide further insight into the
universal nature of the jamming transition and glass transition,
but may also highlight subtle dependencies on the microstructure.

The work of K.W.D was supported by the National Science Foundation
under Grant No. CBET-0853837, and the work of E.R.W. was supported by the National Science Foundation
under Grant No. CMMI-1250235.

%\bibliography{all_ref.bib}

%Merlin.mbs v4.21 2009-07-09.
%

\end{document}